\documentclass{elsart}
\usepackage{natbib}
\usepackage{graphicx,amssymb}
\journal{Advances in Space Research}
\begin{document}
%
\begin{frontmatter}
\title{
Flux Upper Limits of Diffuse TeV Gamma Rays from the Galactic
 Plane Using the Effective Area of the Tibet-II and III Arrays
}
\author[1]{M.Amenomori},
\author[2]{S.Ayabe},
\author[3]{S.W.Cui},
\author[4]{Danzengluobu},
\author[3]{L.K.Ding},
\author[4]{X.H.Ding},
\author[5]{C.F.Feng},
\author[6]{Z.Y.Feng},
\author[7]{X.Y.Gao},
\author[7]{Q.X.Geng},
\author[4]{H.W.Guo},
\author[3]{H.H.He},
\author[5]{M.He},
\author[8]{K.Hibino},
\author[9]{N.Hotta},
\author[4]{Haibing Hu},
\author[3]{H.B.Hu},
\author[10]{J.Huang},
\author[6]{Q.Huang},
\author[11]{M.Izumi},
\author[6]{H.Y.Jia},
\author[11]{F.Kajino},
\author[12]{K.Kasahara},
\author[13]{Y.Katayose},
\author[14]{C.Kato},
\author[10]{K.Kawata},
\author[4]{Labaciren},
\author[15]{G.M.Le},
\author[5]{J.Y.Li},
\author[3]{H.Lu},
\author[3]{S.L.Lu},
\author[4]{X.R.Meng},
\author[2]{K.Mizutani},
\author[7]{J.Mu},
\author[14]{K.Munakata},
\author[16]{A.Nagai},
\author[1]{H.Nanjo},
\author[17]{M.Nishizawa},
\author[10]{M.Ohnishi},
\author[9]{I.Ohta},
\author[2]{H.Onuma},
\author[8]{T.Ouchi},
\author[10]{S.Ozawa},
\author[3]{J.R.Ren},
\author[18]{T.Saito},
\author[11]{M.Sakata},
\author[8]{T.Sasaki},
\author[13]{M.Shibata},
\author[10]{A.Shiomi},
\author[8]{T.Shirai},
\author[19]{H.Sugimoto},
\author[10]{M.Takita},
\author[3]{Y.H.Tan},
\author[8]{N.Tateyama},
\author[8]{S.Torii},
\author[10]{H.Tsuchiya},
\author[10]{S.Udo},
\author[8]{T.Utsugi},
\author[3]{B.S.Wang},
\author[3]{H.Wang},
\author[2]{X.Wang},
\author[5]{Y.G.Wang},
\author[3]{H.R.Wu},
\author[5]{L.Xue},
\author[11]{Y.Yamamoto\corauthref{cor}},
\corauth[cor]{Corresponding author.}
\ead{yamamoto@hep.konan-u.ac.jp}
\author[3]{C.T.Yan},
\author[7]{X.C.Yang},
\author[14]{S.Yasue},
\author[15]{Z.H.Ye},
\author[6]{G.C.Yu},
\author[4]{A.F.Yuan},
\author[10]{T.Yuda},
\author[3]{H.M.Zhang},
\author[3]{J.L.Zhang},
\author[5]{N.J.Zhang},
\author[5]{X.Y.Zhang},
\author[3]{Y.Zhang},
\author[4]{Zhaxisangzhu},
\author[6]{X.X.Zhou}, \\
(Tibet AS$\gamma$ Collaboration)
\address[1]{Department of Physics, Hirosaki University, Hirosaki 036-8561, Japan}
\address[2]{Department of Physics, Saitama University, Saitama 338-8570, Japan }
\address[3]{Institute of High Energy Physics, Chinese Academy of Sciences, Beijing 100039, China }
\address[4]{Department of Mathematics and Physics, Tibet University, Lhasa 850000, China }
\address[5]{Department of Physics, Shandong University, Jinan 250100, China }
\address[6]{Institute of Modern Physics, South West Jiaotong University, Chengdu 610031, China }
\address[7]{Department of Physics, Yunnan University, Kunming 650091, China }
\address[8]{Faculty of Engineering, Kanagawa University, Yokohama 221-8686, Japan}
\address[9]{Faculty of Education, Utsunomiya University, Utsunomiya 321-8505, Japan}
\address[10]{Institute for Cosmic Ray Research, University of Tokyo, Kashiwa 277-8582, Japan }
\address[11]{Department of Physics, Konan University, Kobe 658-8501, Japan}
\address[12]{Faculty of Systems Engineering, Shibaura Institute of Technology, Saitama 330-8570, Japan}
\address[13]{Faculty of Engineering, Yokohama National University, Yokohama 240-8501, Japan }
\address[14]{Department of Physics, Shinshu University, Matsumoto 390-8621, Japan}
\address[15]{Center of Space Science and Application Research, Chinese Academy of Sciences, Beijing 100080, C
hina}
\address[16]{Advanced Media Network Center, Utsunomiya University, Utsunomiya 321-8585, Japan}
\address[17]{National Institute for Informatics, Tokyo 101-8430, Japan}
\address[18]{Tokyo Metropolitan College of Aeronautical Engineering, Tokyo 116-0003, Japan}
\address[19]{Shonan Institute of Technology, Fujisawa 251-8511, Japan}

\begin{abstract}

We obtained new upper limits on the diffuse gamma rays from the inner Galactic (IG) and outer Galactic (OG) planes in 3-10 TeV region, using the Tibet air shower data and new Monte Carlo simulation results. A difference of the effective area of the air-shower array for observing gamma rays and cosmic rays was carefully taken into account in this analysis, resulting in that the flux upper limits of the diffuse TeV gamma rays were reduced by factors of 4.0$\sim$3.7 for 3$\sim$10 TeV than those in our previous results (Amenomori et al., 2002, ApJ, 580, 887). This new result suggests that the inverse power index of the energy spectrum of source electrons responsible for generating diffuse TeV gamma rays through inverse Compton effect should be steeper than 2.2 and 2.1 for IG and OG planes, respectively, with 99\%C.L..
\end{abstract}
\end{frontmatter}
\section{Introduction}

The EGRET observation \citep{Hunter97} of diffuse gamma rays up to 10 GeV from the inner Galactic (IG) and outer Galactic (OG) planes shows a sharp ridge along the Galactic plane.  The flux in the lower energy region ($E<1$ GeV)
is consistent with the conventional calculation \citep[e.g.][]{Dermer86}, although the EGRET flux is higher than the {\sl COS~B} data \citep{Mayer-Ha80} by a factor of about 3 in several GeV. In order to explain the EGRET flux 
in terms of cosmic-ray hadronic interactions with the interstellar matter (ISM), a flatter source proton spectrum with power index of 2.45 \citep{Mori97} or 2.25 \citep{Webber99} is required near the Galactic center with the thickness $|b| \leq 10^\circ$ or $\leq 5^\circ$. On the other hand, \citet{Porter97} suggested that cosmic-ray electrons, in such high energies, may play a significant role in gamma-ray creation.  \citet{Pohl98} showed that if the source electron has a power index  $\beta$ of - 2.0, the EGRET excess above 1 GeV can be well explained by the inverse Compton (IC). Also, \citet{Berezinsky93} calculated the diffuse gamma-ray flux in the wide energy range of TeV-PeV based on the cosmic-ray hadronic interactions with ISM.

In the energy region above 1 TeV, almost all observations have given upper limits on the  flux of diffuse gamma rays from the Galactic plane, except for the recent MILAGRO result giving a 4.5 $\sigma$ excess \citep{Fleysher05} at 1 TeV for 40$^\circ < l < 100^\circ, |b|<5^\circ$, though their inner Galactic plane includes the Cygnus region where some complex features are observed in wavelength bands of radio, infrared and GeV gamma rays. In the previous paper \citep{Amenomori02} we presented the flux upper limits of diffuse gamma rays at 3 TeV and
10 TeV from IG and OG planes using the Tibet-II and Tibet-III data, assuming that the effective area of the array is the same for air showers initiated by gamma rays and cosmic rays with the same primary energy \citep{Amenomori01}.
Since then, the Tibet air shower array has been gradually enlarged and improved by adding more detectors, covering an area of 37000 m$^2$ in which 789 detector are placed at the grid of 7.5 m spacing. We have also done detailed Monte Carlo (MC) simulations to examine the performance of the new air shower array for observing cosmic rays and gamma rays above 1 TeV \citep{Amenomori03}.

 In this paper, we present the results on the flux of diffuse gamma rays obtained with the Tibet air shower array, based on the new MC simulation \citep{Amenomori03} and compare the previous results \citep{Amenomori02}. 

\section{Monte Carlo simulation and data analysis}

A MC simulation for performance of the Tibet array (Fig. 1) is a key subject  in this analysis.  However, we briefly describe its outline because of a page restrictions. 
 In this simulation, primary cosmic rays were  thrown  isotropically into the atmosphere with an appropriate energy spectrum and composition.  In practice,  the heavy dominant (HD) model, given in Fig.~\ref{fig:2}, was  employed, which can well describe the
experimental data in the wide energy rage from 1 TeV to 10$^4$ TeV \citep{Amenomori00, Amenomori01, Amenomori03}. 
The primary gamma rays were assumed to arrive at the Earth with an inverse power spectrum with indices of $\beta$=2.2, 2.4, 2.6, and 2.8 both from the IG plane (20$^\circ \leq l \leq 55^\circ, |b|\leq2^\circ$) and OG plane (140$^\circ \leq l \leq 225^\circ, |b|\leq2^\circ$) \citep{Amenomori02}, as shown later in Fig. \ref{fig:6}.
  
Here, it may be worthwhile to note  that at a high altitude like Yangbajing (606 g/cm$^2$) the size of air showers induced by gamma rays is larger than that induced by protons and other nuclei in the multi-TeV region. The shower size ratio of gamma-ray origin versus proton origin takes a value around 2 $\sim$ 3 for the same primary energy. Hence, the detection efficiency of gamma rays is better than that of protons at Yangbajing.

\begin{figure}[htbp]
  \begin{center}
\begin{minipage}[b]{8.0cm}
    \includegraphics[width=7.5cm]{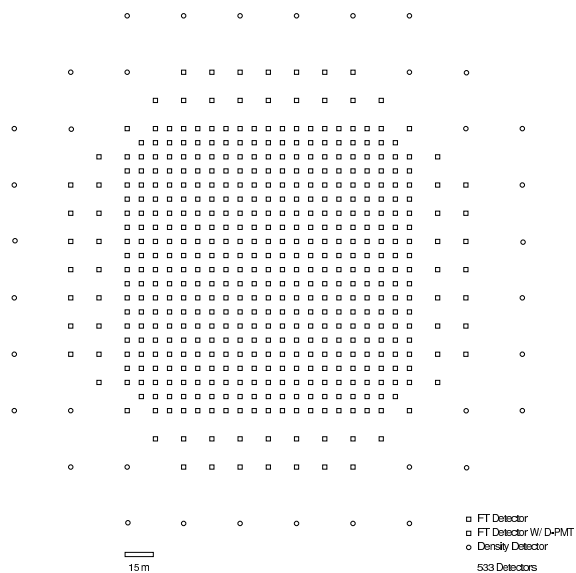}
\end{minipage}
  \end{center}
 \caption{\label{fig:1}  Layout of the Tibet III array in 2000.}
\end{figure}

\begin{figure}[htbp]
  \begin{center}
\hspace*{-12mm}
\begin{minipage}[b]{8.0cm}
    \includegraphics[width=8.5cm]{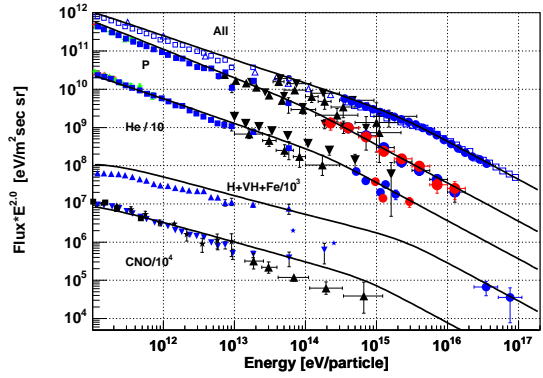}
\end{minipage}
  \end{center}
  \caption{\label{fig:2}  Primary cosmic-ray spectrum in the HD model.}
\end{figure}

\begin{figure}[htbp]
  \begin{center}
\hspace*{4.5mm}
\begin{minipage}[b]{8.0cm}
   \includegraphics[width=7.0cm]{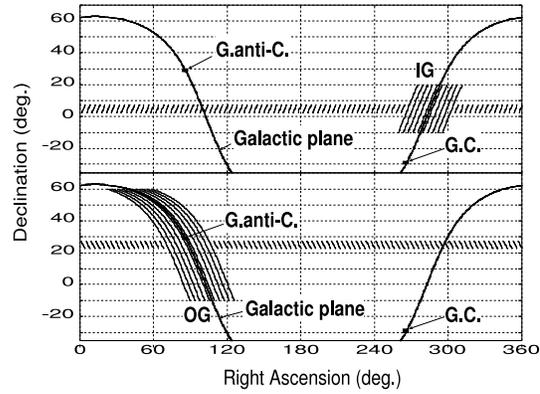}
\end{minipage}
  \end{center}
 \caption{\label{fig:3}  Binning for the IG and OG planes in the analysis.}
 \end{figure}

 Simulated events were generated with the same array configuration as the experiment shown in Fig. \ref{fig:1} and they were analyzed under the same conditions and procedures as the experimental data.  It was confirmed that the mode energies,  $E_{\rm mode}$,  of proton-induced showers observed by the Tibet-III and Tibet-II are 3 TeV and 10 TeV, respectively, where the Tibet-II means a part of 
the Tibet-III overlapping with a detector spacing of 15 m. The Tibet II and III data were analyzed in 4$^\circ$ width binning of the IG and OG planes in the equatorial coordinates as shown in Fig.~\ref{fig:3}, because the respective 120 curved belts are quite the same both in the zenith angle distribution and in the average primary energy of arriving air showers from the IG and OG planes, respectively.

%
%
%
%
%
%
%
%
\begin{figure}[htbp]
  \begin{center}
\begin{minipage}[b]{8.0cm}
   \includegraphics[width=7.5cm]{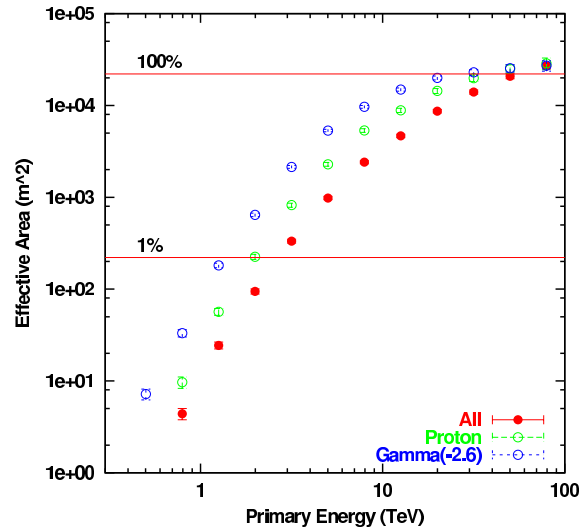}
\end{minipage}
  \end{center}
  \caption{\label{fig:4}  Effective areas of the Tibet III for IG plane.}
 \end{figure}

\begin{figure}[htbp]
 \begin{center}
\begin{minipage}[b]{8.0cm}
  \includegraphics[width=7.5cm]{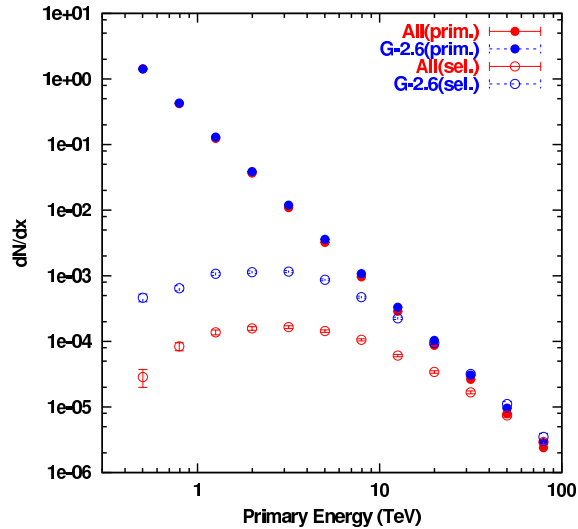}
\end{minipage}
 \end{center}
\caption{\label{fig:5}  Primary spectra and distribution of triggered events for IG plane.}
\end{figure}

\begin{figure}[htbp]
  \begin{center}
\begin{minipage}[b]{8.0cm}
    \includegraphics[width=7.5cm]{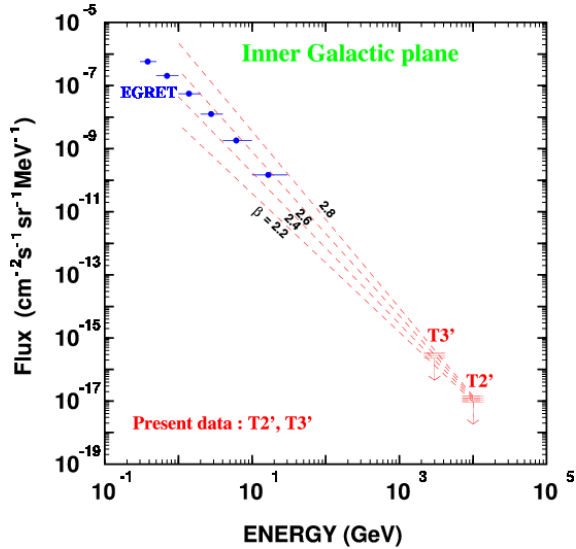}
\end{minipage}
  \end{center}
  \caption{\label{fig:6}  Gamma-ray spectra drawn from 10 TeV.}
 \end{figure}

The effective areas of the Tibet III array for gamma rays, protons and cosmic rays from the direction of IG plane are compared at the given energy bins in Fig.~\ref{fig:4}.  It can be seen that the effective area for gamma rays is about 7 times larger than for the galactic cosmic rays in the multi-TeV region, as already reported elsewhere \citep{Amenomori03}.\\
%
%
 The energy spectra of simulated events coming from the sky region of IG plane, for example, are shown in the differential form in Fig.~\ref{fig:5}, where the spectral index of 2.6 is assumed for both gamma rays and cosmic rays.  
We can see that the mode energy of triggered gamma rays is 1.5$\sim$2.0 times smaller than the galactic cosmic rays for the same trigger condition.  Hence, if this difference is taken into account, the average advantage factor of the effective area for gamma rays is estimated to reduce to 4.0 for $E_{\rm mode}\simeq$3 TeV and 3.7 for 10 TeV in average of the gamma-ray spectral indices of $\beta=2.2\sim$ 2.8.  In the present analysis, the median value $\beta$=2.5 is chosen because of its weak dependence on the spectral index.\\[-1mm]
%
%
%
%
\vspace*{-5mm}
\begin{center}
{\small 
\begin{tabular}{|c|c|c|c|c|} \hline
Array of data taken with & \multicolumn{2}{c|}{Tibet III} & \multicolumn{2}{c|}{Tibet II}\\ \hline 
$E_{\rm mode}$ & \multicolumn{2}{c|}{3 TeV} & \multicolumn{2}{c|}{10 TeV}\\ \hline
Inner or Outer Galactic plane& ~~~I~G~~~ & ~~~O~G~~~ & ~~~I~G~~~ & ~~~O~G~~~ \\ \hline
Significance of excess ($\sigma$)$^\dagger$ & +2.52& +0.25& +1.71 & -0.63\\ \hline
Flux ratio of gamma rays versus cosmic rays$^\dagger$ &&&&\\[-2.5mm]
$I_{\gamma}({\rm at~1~\sigma})/I_{\rm CR}\equiv1/\sqrt{B}~(\times 10^{-4})$&1.95&1.16&2.43&1.45\\ \hline
99\%C.L. upper limit in the original ($\beta$=2.4)$^\dagger$&&&& \\[-2.5mm]
~~$E_{\rm \gamma}^2dN(>E_{\rm \gamma})/dE_{\rm \gamma}$~($\times 10^{-3}$cm$^{-2}$s$^{-1}$sr$^{-1}$MeV)&9.6&3.3&4.0&1.3\\ \hline
Effective area ($\gamma$/CR) ratio of the Tibet array&&&&\\[-2.5mm]
$S_{\rm eff}(\gamma)/S_{\rm eff}$(CR)& {\small\bf 4.0} & {\small\bf 4.0} & {\small\bf 3.7} & {\small\bf 3.7}\\ \hline 
99\%C.L. revised upper limit with $\beta$=2.5&&&& \\[-2.5mm]
~~$E_{\rm \gamma}^2dN(>E_{\rm \gamma})/dE_{\rm \gamma}$~($\times 10^{-3}$cm$^{-2}$s$^{-1}$sr$^{-1}$MeV)&{\bf 2.6}&{\bf 0.88}&{\bf 1.2}&{\bf 0.37}\\ \hline
\end{tabular}
}
\end{center}
\vspace{-3.5mm}
{\small Table~1\\
\hspace*{4mm}Effective area ratios and gamma-ray flux upper limits.}\\
\hspace*{4mm}{\footnotesize The rows with $^\dagger$ are referred to the original paper \citep{Amenomori02}.}\\[8mm]
%
%
It is also found that the effective area ratio gives almost non-difference on the sky region between IG and OG planes. 

\section{Results and discussions}

In this experiment, no significant signal of diffuse TeV gamma rays at the Earth was detected from the Galactic planes.  As in the previous paper \citep{Amenomori02}, the significance ($\sigma$) of the excess events of TeV gamma rays from the IG or OG plane was calculated using a simple formula of $(E-B)/\sqrt{B}$, where $E$ is the number of events on the on-plane and $B$ the number of background events estimated from the neighboring bins around the on-plane. In Table 1, those evaluated values by the previous analysis are tabulated together with the simulated effective area ratio of gamma rays versus cosmic rays and the revised upper limits with a small change of the gamma-ray spectral index $\beta$ from 2.4 to 2.5. In this table, the 3 TeV data were obtained  by the Tibet III array (inner area of 22,050 m$^2$), and the 10 TeV data by the Tibet II array (28,350 m$^2$).

The flux upper limits, thus obtained, may be compared with the EGRET data.  As an example, we show the differential spectrum of diffuse gamma rays from the IG plane in Fig.~\ref{fig:6}. The straight lines denote the interpolation with the assumed spectral indices of 2.2, 2.4, 2.6 and 2.8 at the Earth, drawn from the upper limits at 10 TeV.  From this figure, we may say that the index 2.5 is the most probable value in the GeV-TeV region, if the considerable portion of the EGRET bump in the 1$\sim$10 GeV region is possibly attributed to $\pi^\circ$ decay \citep{Aharonian00}.

\begin{figure}[htbp]
  \begin{center}
\begin{minipage}[b]{8.0cm}
    \includegraphics[width=7.5cm]{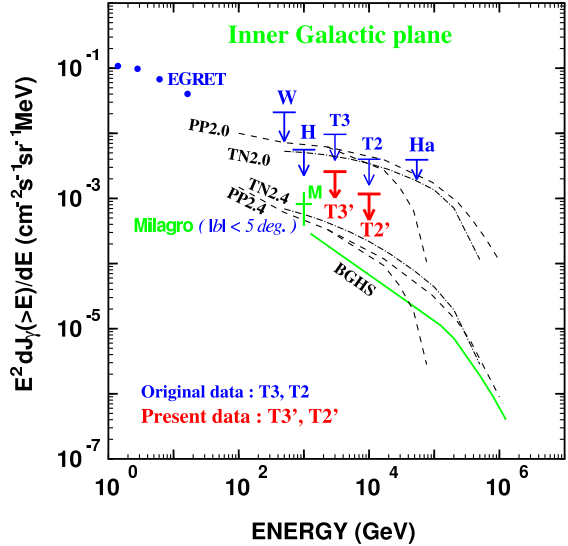}
\end{minipage}
  \end{center}
  \caption{\label{fig:7}  Diffuse gamma rays from the IG plane of 20$^\circ$ 
$\leq l \leq 55^\circ, |b|\leq2^\circ$.}
 \end{figure}

\begin{figure}[htbp]
  \begin{center}
\begin{minipage}[b]{8.0cm}
    \includegraphics[width=7.5cm]{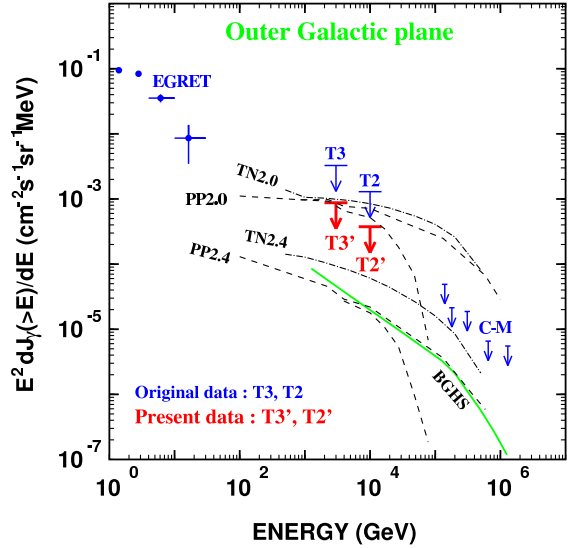}
 \end{minipage}
  \end{center}
\caption{\label{fig:8}  Diffuse gamma rays from the OG plane of 140$^\circ$ 
$\leq l \leq 225^\circ, |b|\leq2^\circ$.}
 \end{figure}

In Figs.~\ref{fig:7} and \ref{fig:8}, we plot the new flux upper limits at $E_{\rm mode}$=3 TeV (Tibet III array) and 10 TeV (Tibet II array) together with the previous ones \citep{Amenomori02}.  New data give more stringent upper limits than the previous ones and this difference can be mostly attributed to a detailed calculation of the effective area of the array for the observation of gamma rays and cosmic rays.  Also, shown in these figures are 
the EGRET data \citep{Hunter97} of $|b| \leq 2^\circ$, the upper limits by Whipple (W) \citep{LeBohec00} with 99.9\%C.L., HEGRA (H) \citep{Aharonian01} with 99\%C.L., where the 2nd and the 3rd experiments scanned a small sky region around {\em l} = 40$^\circ$, and also shown are HEGRA-AIROBICC (Ha) \citep{Aharonian02} and CASA-MIA (C-A) \citep{Borione98} both with 90\%C.L. for comparison.  For the only positive result from MILAGRO (M) \citep{Fleysher05}, it should be noted that the observation of significant excess is located at the Cygnus region rather than in the IG plane according to their significance map.\\

 The calculations for the inverse Compton (IC) were done by \citet{Porter97}(PP2.0 and PP2.4 in figures), and \citet{Tateyama03}(TN2.0 and TN2.4 in figures), where 2.0 and 2.4 are source electron spectral indices. The theoretical curve for gamma rays of hadronic origin through $\pi^\circ$ decay is also given by \citet{Berezinsky93}(BGHS).
%
%
%
%

 In conclusion, our new upper limits with 99\%C.L. can give a constraint that the spectral indices of source electrons for IC should be  steeper than 2.2 in the IG plane and also 2.1 in the OG plane in comparison with the IC theoretical curves, when the power index of 2.5 is adequate for the arriving diffuse gamma rays in the multi-TeV region.

\section{Acknowledgements}

This work is supported in part by Grants-in-Aid for Scientific Research on Priority Areas 712 (MEXT) and by Scientific Research (JSPS) in Japan, and by the Committee of the Natural Science Foundation and by Chinese Academy of Sciences in China.


\end{document}